\documentclass[apj]{emulateapj}
\shorttitle{SN~2008D}
\shortauthors{Bersten et al.}
\begin{document}
\title{Early UV/Optical Emission of The Type Ib SN 2008D} 
\author{Melina C. Bersten\altaffilmark{1}, Masaomi
  Tanaka\altaffilmark{2}, Nozomu Tominaga\altaffilmark{3}, Omar
  G. Benvenuto\altaffilmark{4}\nonumber\footnote{OGB is member of
    the Carrera del Investigador Cient\'{\i}fico de la Comisi\'on de
    Investigaciones Cient\'{\i}ficas  
de la Provincia de Buenos Aires (CIC), Argentina.}, Ken'ichi Nomoto\altaffilmark{1}}   

\affil{\altaffilmark{1} Kavli Institute for the Physics and Mathematics of
  the Universe, Todai Institutes for Advanced Study, University of
  Tokyo, 5-1-5 Kashiwanoha, Kashiwa, Chiba 277-8583, Japan} 
\affil{\altaffilmark{2} National Astronomical Observatory of Japan, 
 2-21-1 Osawa, Mitaka, Tokyo 181-8588, Japan}

\affil{\altaffilmark{3} Department of Physics, Faculty of Science and
  Engineering, Konan University, 8-9-1 Okamoto, Kobe, Hyogo 658-8501,
  Japan}  
\affil{\altaffilmark{4} Facultad de Ciencias Astron\'omicas y
  Geof\'{\i}sicas, Universidad Nacional de La Plata, Paseo del Bosque
  S/N, B1900FWA La Plata, Argentina}  

\email{melina.bersten@ipmu.jp}

\submitted{Submitted to ApJ on October 25, 2012 --- Accepted on
  February 26, 2013.} 

\begin{abstract}
\noindent 
We propose an alternative explanation for the post-breakout emission
of SN~2008D associated with the X-ray transient 080109. Observations
of this object show a very small contrast of $0.35$ dex between the
light-curve minimum occurring soon after the breakout, and the main
luminosity peak that is due to radioactive heating of the
ejecta. Hydrodynamical models show that the cooling of a shocked
Wolf-Rayet star leads to a much greater difference ($\gtrsim 0.9$
dex). Our proposed scenario
is that of a jet produced during the explosion which deposits
$^{56}$Ni-rich material in the outer layers of the ejecta. The
presence of high-velocity radioactive material allows us to reproduce the
complete luminosity evolution of the object. Without outer $^{56}$Ni
it could be possible to reproduce the early 
emission purely from cooling of the shocked envelope by assuming a
larger progenitor than a Wolf-Rayet star, but that would require an
initial density structure significantly different from what is
predicted by stellar evolution models. 
Analytic models of the cooling phase have been proposed reproduce
the early emission of SN~2008D with an extended progenitor. However,
we found that the models are valid only until 1.5 days after the explosion 
where only two data of SN~2008D are available.
We also discuss the possibility of the interaction of the ejecta with a
binary companion, based on published analytic expressions. However,
  the binary separation required to fit the early emission should be
  $\lesssim 3 \,R_\odot$, which is too small for a
system containing two massive stars.
\end{abstract}
  

\section{INTRODUCTION}
\label{sec:intro}
  Supernova (SN) 2008D attracted a good deal of attention because of
  its unusual observational characteristics. Most outstandingly, the
  serendipitous detection of the X-ray transient (XRT) 080109
  associated with the SN explosion, during the Swift follow-up of
  SN~2007uy, another SN in the same galaxy, NGC~2770 
\citep{2008GCN..7159....1B,kong08_atel}. The optical counterpart of
the XRT was revealed at the same position of the transient a few hours
after, allowing to have unprecedented early coverage of the SN
emission. Initially the SN  was classified as a broad-line Type Ic SN
(SN~Ic) \citep{2008CBET.1205....1B,2008CBET.1205....2V} based on the
broad absorptions and lack of hydrogen and helium in the first
spectrum. Later on, however, the spectra revealed the presence of
strong He I lines, which changed the classification to Type Ib
\citep{2008CBET.1222A...1M,2009ApJ...692L..84M}. 

Type Ib SNe, as well as Type Ic and the transitional Type IIb
  SNe, are believed to be the
result of the core collapse of massive stars ($M_{\mathrm{ZAMS}} \gtrsim 10$
$M_\odot$) that have lost most or all of their hydrogen (and often
helium) layers before the explosion. For that reason, they are  called
``stripped-envelope SNe'' \citep{1996ApJ...462..462C}. However, the
mechanism by which the envelope is removed is not fully
understood. Strong winds of 
single massive progenitors ($M \gtrsim 30$ $M_\odot$), sudden eruptions and
binary interaction have been proposed as possible explanations
\citep[][among
  others]{1994A&A...287..803M,1995PhR...256..173N,2004ApJ...616..525O},
with the binary origin being the most probable scenario. In all cases
a Wolf-Rayet (WR) structure is expected before the explosion, although
for SNe~IIb a thin H envelope is expected to remain before the
explosion (e.g., see SN~1993J
\citep{1993Natur.364..507N,1993Natur.364..509P}, and SN~2011dh
\citep{2012arXiv1207.5975B}). Interest in 
these objects has recently grown due to its 
connection with long gamma-ray bursts (GRB). To date six SNe have been
associated with 
GRB, one of which also showed XRT. In all of these cases a highly
energetic Type Ic SN, also called ``hypernova'' was observed. Note,
however, that the term hypernova is more generally used, and refers to
SNe with high luminosity or unusually broad lines, independently of
their association with GRB. The origin of hypernovae is thought to be
a rapidly rotating and accreting compact object
\citep{2001ApJ...550..410M}, or a magnetar
\citep{2004ApJ...611..380T}, which should produce relativistic
outflows.

The nature of the XRT associated with SN~2008D is controversial. Some
authors are in favor of the supernova shock-breakout origin
\citep{2008Natur.453..469S,2008ApJ...683L.135C}, while others consider
that the transient was caused by a mildly relativistic jet penetrating
through the envelope of the progenitor star
\citep{2008Sci...321.1185M,2008MNRAS.388..603L,2008cosp...37.3512X}. In
the latter case, the XRT could represent a transition between the most
energetic hypernovae and standard core-collapse SNe. However, the
presence of a GRB or the radio detection of superluminal motions
caused by a long-lived relativistic outflow were firmly ruled out for
SN~2008D
\citep{2008Natur.453..469S,2009ApJ...694L...6B}. Nevertheless, the
lack of confirmation of a thermal component for the
XRT \citep{2009ApJ...702..226M}, and the 
strong evidence of an asymmetric explosion
\citep{2009ApJ...702..226M,2010A&A...522A..14G}, possibly bipolar
\citep{2009ApJ...700.1680T,2009ApJ...705.1139M}, leaves room for the
mildly relativistic-jet scenario.

Another interesting feature of of SN~2008D was
its double-peaked optical light curves (LC). The first peak occurred
at $\approx$ 1 day after the XRT, and
was similar to that of SN~2006aj, which was connected with a GRB
\citep{2006Natur.442.1008C}. The second and main peak happened at
$\approx$ 20 days after the XRT, consistently with other SNe~Ib and Ic, and
its origin is related with the decay of radioactive material
synthesized during the explosion. By modeling the emission
around the main peak, explosion parameters of SN~2008D, such as ejecta mass
($M_{\mathrm{ej}}$), kinetic energy ($E_{K}$), and nickel mass
($M(^{56}\mathrm{Ni})$) were estimated in previous studies. Using
analytic models, \citet{2008Natur.453..469S} found  $M_{\mathrm{ej}}=
3-5$ $M_\odot$, and $E_{K}= 2-4$ foe (1 foe = $1 \times 10^{51}$ erg
s$^{-1}$). Based on Monte Carlo simulations, 
\citet{2008Sci...321.1185M} suggested that $M_{\mathrm{ej}}=7$
$M_\odot$, $E_{K}=6$ foe, and $M(^{56}\mathrm{Ni})= 0.09$
$M_\odot$. \citet{2009ApJ...692.1131T} (T09 hereafter) subsequently
presented an exhaustive analysis of SN~2008D using a set of progenitor
models and hydrodynamics/nucleosynthesis calculations. They found a
very good agreement with the LC and the spectra for a model with
$M_{\mathrm{ej}}= 5.3 \pm 1$ $M_\odot$, $E_{K}=6
\pm 2.5$ foe, and $M(^{56}\mathrm{Ni})= 0.07$ $M_\odot$.

Early UV/optical emission is expected to occur after the arrival of
 the shock wave at the surface of the progenitor (shock break-out) and
 before the re-brightening due to the decay of radioactive material. This
 emission is a consequence of the nearly adiabatic cooling due to the
 expansion of the outermost layers of 
the ejecta. The observations of the shock break-out and the following
emission provide very valuable information about the structure of the
star previous to the explosion. The duration of this early phase
depends strongly on the size 
of the progenitor. For compact Wolf-Rayet stars, which are proposed
progenitors of SNe~Ib and Ic, a duration of a few
days is expected. Consequently, 
catching a SN during this phase is quite a challenge and only for a
handful of objects has this been possible. 

Naturally, the origin of the earliest part of the LC of 
SN~2008D has been associated in the literature to the adiabatic
cooling of the outer layers of the ejecta 
\citep{2008Natur.453..469S,2009ApJ...702..226M,2011ApJ...728...63R}. However,  
different authors arrived at different conclusions 
regarding the progenitor radius. Some suggested a value of
$\approx$ $1 \, R_\odot$ \citep{2008Natur.453..469S,2011ApJ...728...63R}
while others proposed a larger value of $\approx$ $9 \, R_\odot$
\citep{2009ApJ...702..226M,2008ApJ...683L.135C}. In all these cases the
estimations were based on analytic models for the early
emission. Note that \citet{2008Sci...321.1185M} and T09 performed numerical 
simulations, but they did not attempt a hydrodynamical modeling
of the early emission so as to help discriminate between both
possibilities.

 Given the unique information about the progenitor structure that
  is provided by the early emission we have performed new
  hydrodynamical calculations for SN~2008D, focusing on 
    this phase.  
Surprisingly, we found that our hydrodynamical models are not
consistent with the cooling-phase explanation given
in the literature for 
this object. Hydrodynamical models predict a much larger contrast between the
luminosity minimum which occurs after the breakout, and the  
 luminosity peak due to radioactive decay than what was 
observed for SN~2008D. This has also been noted recently by
\citet{2011MNRAS.414.2985D} using a set of hydrodynamical and spectral
calculations of SNe~Ib and Ic. 

Here we present the first radiation-hydrodynamical models for the early
    emission of SN~2008D and propose an alternative explanation
for this emission based on a double-peaked $^{56}$Ni
distribution. The data and  hydrodynamic 
code employed are described in \S~\ref{sec:datamodel}. A comparison
with previous hydrodynamical results is shown in 
\S~\ref{sec:CT09}. Our proposed model for SN~2008D is presented in
\S~\ref{sec:optimal}. In \S~\ref{sec:discussion} we compare our model
with the analytic models used in the literature, and discuss 
alternative explanations for the early emission. Finally, our
conclusions are presented in \S~\ref{sec:conclusion}.

\section{DATA AND MODEL}
\label{sec:datamodel}
\subsection{Observational Material}
The bolometric LC (Lbol) of SN~2008D was calculated by
\citet{2009ApJ...702..226M}  
using $UVW1BVRr'Ii'JHK_s$ broad-band photometry for $t<31$ days, referred
to the onset of the XRT, i.e., $\mathrm{JD} = 2454475.06$ (in what
follows, all times will be referred to this moment).  Here we
  adopt the estimates of Lbol from black-body fits to the
  broadband photometry, which according to \citet{2009ApJ...702..226M}
  are more accurate at early times than direct integration of the
  observed flux. 
At later times we include bolometric calculations provided by T09 (see
their Appendix~1) using optical and near-infrared (NIR)
data obtained with the MAGNUM telescope and the Himalayan Chandra
telescope. Also available in the literature are 
two earlier data points, one at  $0.14$ days observed with the  
Swift telescope in the $UVW2$, $UVW1$ and $U$ filters
\citep{2008Natur.453..469S}, and another at $0.44$ days  
in the $BVRI$ bands \citep{2008Sci...321.1185M}. The integrated
flux for each of these observations was calculated respectively by the
authors. Here we adopt, for $t=0.44$ days, 
the sum of the luminosity in the $BVRI$ bands, plus the contribution
of the UV as estimated from the earlier Swift observations. 
The uncertainty of this point was assumed to be the sum in quadrature of
the uncertainties of both contributions. 
We denote this earliest point with a different symbol (square) in all
figures where it is included in order to indicate its different
origin.  The distance and total reddening assumed in
the calculations are $d= 32$ Mpc, and $E(B-V) = 0.65$ mag
\citep{2008Sci...321.1185M}.  

The bolometric LC reveals a double-peaked shape that is also observed
in the broad-band photometry 
\citep{2008Natur.453..469S,2008Sci...321.1185M}. We
assume throughout this paper that the earliest data point is confident and
therefore the early peak shape of the LC is robust. The first peak
occurred at about 1 day with a luminosity of $L= 1.\, \times \,
10^{42}$ erg s$^{-1}$, and the second or main maximum happened at $\approx$19 
days with a luminosity of $L= 1.62 \, \times \, 10^{42}$ erg s$^{-1}$,
corresponding to $M_{\mathrm{bol}}= -16.8$ mag. Therefore, SN~2008D
had a normal peak luminosity, and a rise time at the long end of the
observed range for SNe~Ib and Ic
\citep{2006AJ....131.2233R,2012ApJ...753..180D}. The contrast of
  luminosity between the main peak and the dip that occurs before
  the re-brightening of the LC due to radioactive material is only 
  of $0.35$ dex. 

Apart from the bolometric LC, photospheric velocities and color
temperatures are also compared with our models. We employ He-line
and photospheric velocities derived from the spectral modeling by T09,
and color temperatures calculated by \citet{2009ApJ...702..226M}. 

\subsection{Light Curve Models}
\label{code}
Synthetic LCs were calculated using the spherical, Lagrangian, LTE,
hydrodynamical code described by \citet{2011ApJ...729...61B} (BBH11,
hereafter). The code 
solves radiation transport in the 
flux-limited diffusion approximation, including $\gamma$-ray
transfer in gray approximation. Any distribution of the radioactive
material is allowed and the energy deposition due to radioactive
  decay is computed in each layer of the entire ejecta. The
ionization structure is determined 
by solving the Saha equation taking into account the most relevant
elements in the progenitor structure. The Rosseland mean opacity is 
calculated using OPAL tables \citep[][and references
  therein]{1996ApJ...464..943I}, and the empirical
relation used in T09, which is based on electron scattering
opacity as derived from the TOPS database
\citep{1995ASPC...78...51M,2005ApJ...624..898D}, and including effects
of lines. The explosion is simulated by
injecting near the center of the object a certain amount of energy, in
a thermal form, during a short interval. The code does not 
explicitly solve for the explosive nucleosynthesis produced
during the shock propagation, but it implicitly takes this into
consideration in the chemical composition assumed for our initial
models. 

As pre-supernovae models, stellar evolution calculations 
of He stars by \citet{NH88} were adopted. T09 studied five
different He star models with masses of 4, 6, 8, 10, and 16 $M_\odot$,
using a variety of explosion parameters (see their Table~1). They
found that the models with He mass of 6 and 8 $M_\odot$ (He6 and He8,
respectively) were most consistent with the spectra and LCs
of SN~2008D. They derived a kinetic energy of $E_{K} = (6.0 \pm 2.5)
\, \times \, 10^{51}$ erg, an ejecta mass of $M_{ej}= 5.3 \pm 1.0$
$M_\odot$, a $^{56}$Ni mass of $M(^{56}\mathrm{Ni})= 0.07$ $M_\odot$, and 
 a compact-remnant mass in the range of $1.6 - 1.8$ $M_\odot$. 
 
We have based the current analysis on the results of T09, and so we
calculated the hydrodynamics for these two optimal 
models (He6 and He8), which correspond to main-sequence masses
of 20 and 25 $M_\odot$, as derived from the $M_{\mathrm{MS}} - M_\alpha$   
relation of \citet{1980SSRv...25..155S}. The chemical abundance
distribution left by the explosive nucleosynthesis is assumed as a
pre-explosion condition (see T09 for more details).  One characteristic
of these nucleosynthesis calculations in spherical symmetry is that
$^{56}$Ni is confined in the innermost layers of the ejecta, which
makes it very difficult to explain the timescale of the rising
part of the LC normally observed for SNe~Ib and Ic. Instead, a more
extensive mixing of $^{56}$Ni, associated to multi-dimensional effects,
was used by T09, and we also adopted it here (see \S~\ref{sec:CT09}
for more details). Note that  mixing of  $^{56}$Ni to large radius
of the ejecta of CCSNe has been successfully produced in recent  3D
numerical calculations   
\citep{2010ApJ...714.1371H,2010ApJ...723..353J}. In addition,
\citet{2012MNRAS.424.2139D} have claimed that all SNe classified as Ib
require efficient mixing of $^{56}$Ni in the helium-rich layers.

\section{HYDRODYNAMIC MODEL OF SN~2008D}
\label{sec:Hydro}
We present here our hydrodynamic calculations for SN~2008D using the He6
and He8 initial models. In \S~\ref{sec:CT09} we compare our results
with those of T09 by focusing on the second peak of the bolometric LC
($t>5$ days), which determines the global properties of the  
SN. The early emission is analyzed in \S~\ref{sec:optimal}, where we
present a model that consistently reproduces the the first and second
peaks of the LC. 

\subsection{Comparison with T09}
\label{sec:CT09}

As it is well known, the shape of the LC depends on the explosion
kinetic energy $E_K$, the ejected mass $M_{\mathrm{ej}}$, and the mass and
distribution of $^{56}$Ni. Analytic expressions for these dependencies
were given by \citet{1982ApJ...253..785A}, where the width of the LC
peak is $\tau_{\mathrm{LC}} \propto$ $M_{\mathrm{e}j}^{3/4} \, E_K^{-1/4}$, and
the peak luminosity is $L_{\mathrm{peak}} \propto M_{\mathrm{Ni}} \,
\tau_{\mathrm{LC}}^{-1}$. Therefore, it is possible to estimate 
explosion parameters by comparing models and observed bolometric
LCs. However, various combinations of $M_{\mathrm{ej}}$ and $E_K$ can fit the
LC, and spectra modeling is needed in order to break the degeneracy of
the parameters. This type of analysis was done by T09 and they found a 
very good agreement between models and observations using the He6 and
He8 initial models with the following explosion parameters: for He6, $E_K =
3.7$ foe, a cut mass of $M_{\mathrm{cut}} = 1.6 \, M_\odot$, and
$M_{\mathrm{Ni}}= 0.065 \, M_\odot$; and for He8, $E_K = 8.4$ foe,
$M_{\mathrm{cut}}= 1.8 \, M_\odot$, and $M_{\mathrm {Ni}} = 0.07 \,
M_\odot$. A constant $^{56}$Ni distribution up to 7000 (9000) km
s$^{-1}$ for model He6 (He8) was assumed to account for the rise time
to the main peak. 

\begin{figure}
\begin{center}
\includegraphics[scale=.33,angle=-90]{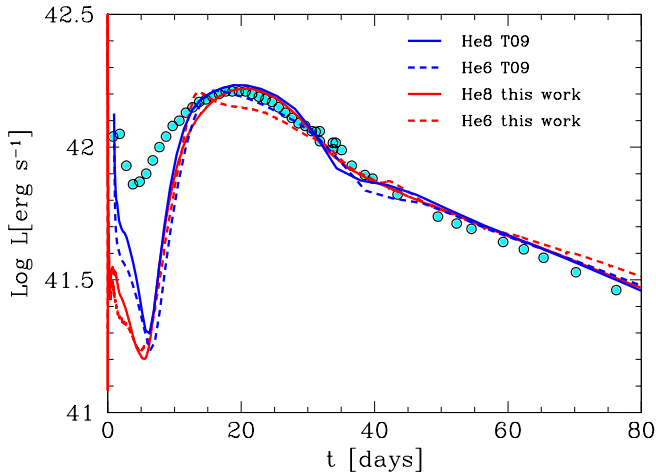}
\caption{Comparison between the bolometric LC for models He6 (blue lines) and
 He8 (red lines) calculated in this work (solid lines) and by T09
 (dashed  lines). Both works adopted the same
 progenitor models and physical parameters. The
 differences are only related with the hydrodynamic code used (see
 \S~\ref{sec:CT09} for more details). The observed bolometric LC of
 SN~2008D is also included (cyan dots). 
\label{fig:LC1}}
\end{center}
\end{figure}

\begin{figure}
\begin{center}
\includegraphics[scale=.40]{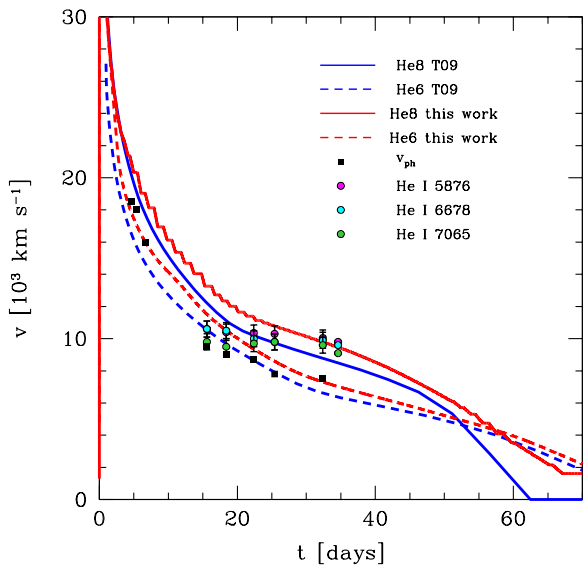}
\caption{Evolution of the photospheric velocity for models He6 (blue lines) and
  He8 (red lines) calculated in this paper (solid lines) and by T09
  (dashed lines). For comparison, observed He line velocities (circle
  dots) and photospheric velocities estimated through spectra modeling
  by T09 are also shown.\label{fig:velo}}  
\end{center}
\end{figure}

We have used the code of BBH11 to calculate
bolometric LCs and photospheric evolution for the same models (He6 and
He8) and explosion parameters as those of T09. As opposed to T09, our
calculations solve the hydrodynamics coupled to the radiative
transfer, allowing us to model consistently the earliest phases of the
SN evolution. Figure~\ref{fig:LC1} shows a comparison of
our results with those of T09, along with the observed bolometric LC
of SN~2008D. The photospheric velocities are compared in  
Figure~\ref{fig:velo} where we include the He lines velocities as well as the 
photospheric velocity estimated from spectra modeling by T09. From
these figures we see that the agreement between both 
 models is reasonable, considering the differences in the calculation
methods. The LCs look remarkably similar for times $t \gtrsim 5$ days.
It is not striking that the largest differences appear at the earliest
epochs, since the code in BBH11 self-consistently calculates the shock
wave propagation, the breakout and the later evolution, while
in the calculations of T09, the hydrodynamics and the radiative
transport were computed with different codes, switching from one
code to other when the homologous expansion was achieved. There is a
small systematic difference between the photospheric velocities of
both models. This is probably related to slight differences in the
re-gridding of the initial model, which produces a small difference in
the actual total mass of the progenitor (of $\approx 0.4\, M_\odot$).
begin{figure}

Both our calculations and those of T09 provide very good matches to
the observations around the main peak. Note that
model He8 is the one that best represents the bolometric LC, while model
He6 gives a better fit to the velocities, specially the photospheric
velocities. Some intermediate model between these two seems to be the
most plausible progenitor, as suggested in T09. However, at times
before 8 days,  
the models predict much lower luminosity than the observed one. As
described in \S~\ref{sec:optimal}, this can be overcome by including a
small amount of $^{56}$Ni in the outermost layers of the ejecta. 

\subsection{A High-Velocity blob of $^{56}$Ni}
\label{sec:optimal}
 It is clear from Figure~\ref{fig:LC1} that the models presented in
\S~\ref{sec:CT09} cannot explain the first peak shown by the
observations. The difference in luminosity between the
hydrodynamical models and the observations at early times are larger
than $0.5$ dex, which is much greater than any possible uncertainty related
with the calculations and observations. The models predict a
brightness contrast between the dip and the main peak that is
larger than $0.9$ dex, while the observations show a contrast of
only $0.35$ dex. Although the early behavior of SN~2008D was
attributed in the literature to the cooling of the outer envelope of
the ejecta after being heated by the shock wave, our calculation
suggests that this cooling happened very quickly, and
in less than 1 day most of the energy deposited by the shock was
degraded. Therefore, unless we assume a different structure for the
progenitor ---i.e., lager radius and/or more massive
external envelope---, some source of extra energy in the outer layers
of the ejecta is required in order to reproduce the first peak of the
LC. Here we study this possibility and leave the
discussion of other alternatives for \S~\ref{sec:discussion}.

We artificially placed some $^{56}$Ni in the outer layers of the
ejecta as the source of extra energy needed to explain the first
peak.  This material may have been carried by a jet-like phenomenon
produced during an aspherical explosion. The presence
of a jet was proposed by \citet{2008Sci...321.1185M} to explain
the XRT associated with SN~2008D, as well as the broad-line appearance
of the spectrum at $t \lesssim 3-5$ days. The double-peaked oxygen
lines detected in late spectra
\citep{2009ApJ...692L..84M,2009ApJ...702..226M,2009ApJ...700.1680T} 
provide additional evidence of the asphericity during the explosion. A 
spectropolarimetry study of SN~2008D by \citet{2009ApJ...705.1139M}
suggest that a jet was produced but that it stalled in the C+O
core. Maund et~al. base this conclusion mainly on the observed low
degree of \ion{O}{1} line polarization. However, for this SN the
\ion{O}{1} $\lambda$ 7774 line is so weak that one cannot expect any
strong polarization associated with it. Weak lines do not produce high
polarization, as shown, for example, in \citet{2012ApJ...754...63T}. 

A double-peaked $^{56}$Ni distribution was previously
suggested to model the double-peaked light curve of SN~2005bf
\citep{2005ApJ...633L..97T,2006ApJ...641.1039F}. The presence of an
unobserved jet in SN~2005bf was speculated as being responsible
for the anomalous $^{56}$Ni distribution.  Despite the overall observational
differences between SN~2005bf and SN~2008D, some similarities in the
polarization properties of both objects were found
\citep{2009ApJ...705.1139M} that could indicate similarities in the
explosion geometry. 

\begin{figure}
\includegraphics[scale=.33, angle=-90]{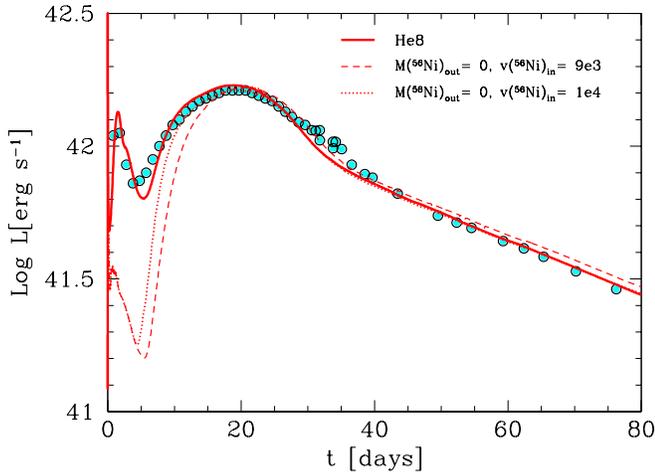}
\caption{Bolometric LC of our optimal model (solid line) compared with
  the observations of SN~2008D. This model can consistently reproduce
  the two observed peaks. For comparison, two models without
  external $^{56}$Ni and slightly different internal $^{56}$Ni
  distributions are shown. A model with $^{56}$Ni mixed up to
  9000 km s$^{-1}$ as assumed in T09 is shown in dashed line, and a
  model with the same internal $^{56}$Ni distribution used in the optimal
 model, i.e. up to $\approx$10000 km s$^{-1}$ is shown in dotted line.  
 Note that both models without the external $^{56}$Ni show a difference
 with the observations larger than $0.5$ dex at early
 times. \label{fig:model}}  
\end{figure}

In Figure~\ref{fig:model} we show the bolometric LC for our
model with $^{56}$Ni in the outer layers (solid line) compared with
the observations. This ``optimal model'' is similar to the He8 model
presented in \S~\ref{sec:CT09} (dashed line in the figure) but  with
an extra amount of $0.01 \, M_\odot$ of $^{56}$Ni in the outermost
layers of the ejecta (at $v > 20,000$ km s$^{-1}$) added to explain
the early emission at $t \lesssim 5$ days. In addition, the optimal
model has a slightly extended distribution of internal $^{56}$Ni up to
$\approx$10000 km s$^{-1}$ (as compared with 9000 km s$^{-1}$ that was
used in the model of T09) to improve the agreement with the data
during the rise to the main peak. For clarity, in
Figure~\ref{fig:model} we show two models that lack external
$^{56}$Ni, with internal distributions up to 9000 and 10000 km
s$^{-1}$ (dashed and dotted lines, respectively). In the rest of this
paper the model without external $^{56}$Ni will be the latter one and
will be referred to as He8.

The agreement between the optimal model and the
observations is excellent. The assumption of external $^{56}$Ni
allowed us to reproduce the first and second peaks consistently. 
The luminosity contrast between dip and main peak is now similar
to that shown by the observations. 

\begin{figure}
\begin{center}
\includegraphics[scale=.3]{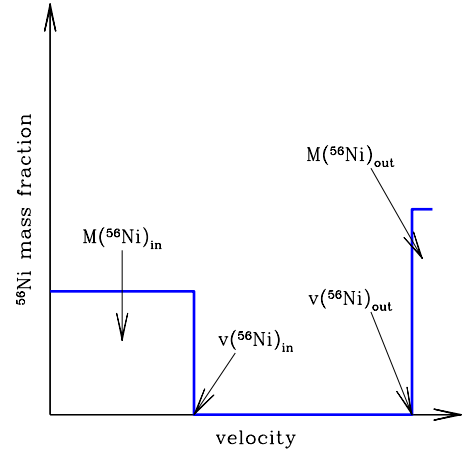}
\caption{Schematic doubly-peaked $^{56}$Ni distribution characterized
  by the following parameters: the mass of internal $^{56}$Ni,
  $M(^{56}\mathrm{Ni})_{\mathrm{in}}$, the mass of external $^{56}$Ni, 
  $M(^{56}\mathrm{Ni})_{\mathrm{out}}$, the minimum velocity for the
  external $^{56}$Ni, $v(^{56}\mathrm{Ni})_{\mathrm{out}}$ and the
  maximum velocity for the internal $^{56}$Ni,
  $v(^{56}\mathrm{Ni})_{\mathrm{in}}$. \label{fig:Nidist}}
 \end{center}
\end{figure}

\begin{figure}
\begin{center}
\includegraphics[scale=.33, angle=-90]{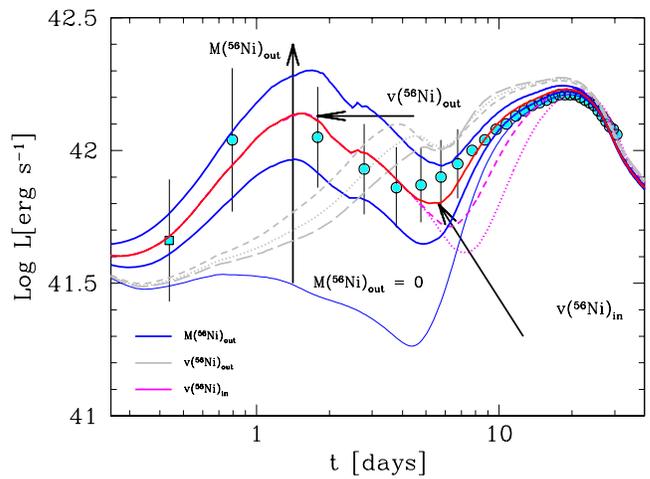}
\caption{Sensitivity of the early LC on the distribution of
  $^{56}$Ni. Three different values of
  $M(^{56}\mathrm{Ni})_{\mathrm{out}}$,
  $v(^{56}\mathrm{Ni})_{\mathrm{out}}$ and
  $v(^{56}\mathrm{Ni})_{\mathrm{in}}$ have been used (see
  Figure~\ref{fig:Nidist}). The arrows indicate the change produced in
  the LC by the increase of each parameter. A model without $^{56}$Ni
  is shown for comparison. Time is plotted in logarithmic scale to
  show more clearly the behavior of the early LC.\label{fig:LCNi}}  
 \end{center}
\end{figure}

In order to obtain our optimal model, several $^{56}$Ni distributions
were explored. In Figure~\ref{fig:Nidist} we schematically show a
double-peaked $^{56}$Ni distribution where the internal and external
components were assumed to be step-like functions within some internal
and external velocities. This distribution can be characterized with
three parameters: the mass of external $^{56}$Ni,
$M(^{56}\mathrm{Ni})_{\mathrm{out}}$, the minimum velocity for the
external $^{56}$Ni, $v(^{56}\mathrm{Ni})_{\mathrm{out}}$, and the
maximum velocity for the internal $^{56}$Ni,
$v(^{56}\mathrm{Ni})_{\mathrm{in}}$. The effect of the variation of
these parameters on the early LC is shown in Figure~\ref{fig:LCNi}. We
see that (1) larger $M(^{56}\mathrm{Ni})_{\mathrm{out}}$ produces a
more luminous first peak, (2) higher
$v(^{56}\mathrm{Ni})_{\mathrm{out}}$ produces an earlier first peak, 
and (3) lower $v(^{56}\mathrm{Ni})_{\mathrm{in}}$ translates
to deeper and later minimum, or later rise to the second
peak. Specifically, we have adopted the following values
 for our optimal model: $M(^{56}\mathrm{Ni})_{\mathrm{out}}= 0.01 M_\odot$,
  $v(^{56}\mathrm{Ni})_{\mathrm{out}}= 22000$ km s$^{-1}$, and
   $v(^{56}\mathrm{Ni})_{\mathrm{in}}= 10000$ km s$^{-1}$.

Although we do not provide a detailed calculation to account for the
occurrence of such $^{56}$Ni distribution, it is remarkable that it
allows us to obtain a very good fit to the observation and also to
explore its effect on the early LC. We also explored the possibility of using
smoother functions for the $^{56}$Ni distribution, but we found that
they failed to reproduce the two peaks shown in the observations, and
instead they tended to produce a plateau-like shape. Given the large
uncertainties in the early observations, we emphasize that the
critical feature of our model is to have some $^{56}$Ni in the outer
layers but not the exact amount and shape of its distribution. 

\begin{figure}
\begin{center}
\includegraphics[scale=.33, angle=-90]{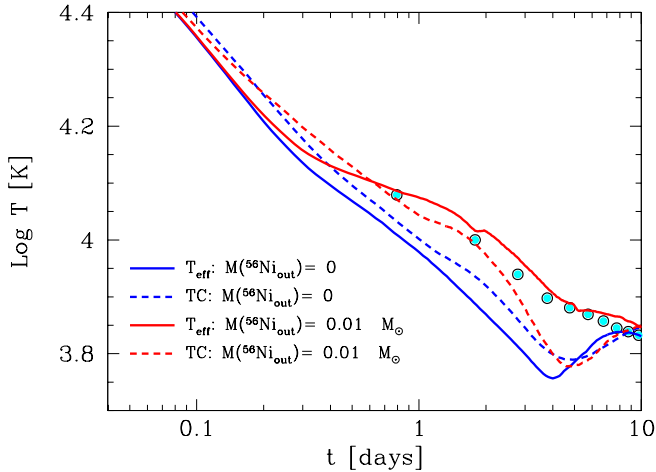}
\caption{Color (dashed lines) and effective temperature (solid lines)
  evolution for our optimal model (red) and for the same model
  without external $^{56}$Ni (blue). The color temperature of SN~2008D
  calculated by \citet{2009ApJ...702..226M} using broad-band
  photometry is shown for comparison (dots). \label{fig:Teff}}  
\end{center}
\end{figure}

Another important parameter to compare with the observations is the
temperature. Figure~\ref{fig:Teff} shows the effective (solid
lines) and color (dashed lines) temperatures for the optimal model
(red) and the model without external $^{56}$Ni (blue),
compared with the observed color temperature of SN~2008D
\citep{2009ApJ...702..226M}.
 The color temperature ($T_C$) gives information about the continuum
 spectral energy distribution. To calculate $T_C$ from the models it is
needed to know the layer in the ejecta where the spectrum is formed. An
approximate estimation of $T_C$ is given by the temperature at the
``thermalization'' depth \citep{1992ApJ...393..742E}, below which the
gas and radiation field are in
equilibrium. Following \citet{1992ApJ...393..742E}, we calculated the
``thermalization'' depth as the layer where $3 \,\tau_{\mathrm{abs}}
\, \tau_{\mathrm{sct}} \approx 1$. Here $\tau_{\mathrm{sct}}$ is the
optical depth for scattering and $\tau_{\mathrm{abs}}$ the optical
depth for absorption determined using $\kappa_{\mathrm{abs}}= \kappa -
\kappa_{\mathrm{sct}}$ where $\kappa$ is the Rosseland mean
opacity calculated with OPAL tables, and $\kappa_{\mathrm{sct}}$ is
the scattering opacity calculated by solving the Saha equations. The
effective temperature is defined as $T_{\mathrm{eff}}^4= L / (4 \pi \,
\sigma R_{\mathrm{ph}}^2)$, where $R_{\mathrm{ph}}$ is the radial coordinate
at the photosphere position defined as the layer where $\tau=1$ and
$L$ is the radiative luminosity plus the luminosity of $^{56}$Ni
decay deposited above the photosphere\footnote{As stated in
  Section~\ref{code}, we have solved the gamma-ray transfer which allow us to
compute the actual  deposition of energy of gamma-ray in the entire
  ejecta (in particular, above the photosphere). Such deposition can be
  appreciably lower than the total available radioactive energy
  release, specially when the envelope 
  becomes very dilute.}. From Figure~\ref{fig:Teff} we
see, as expected, that the model without external $^{56}$Ni has
higher color than effective temperature, but neither of them provides
a satisfactory agreement with the data. On the other hand, the
temperatures of the optimal model compare acceptably well with the
data, but unlike what is expected, the effective temperature is higher
than the color temperature for $t\gtrsim 1$ day. This can be 
understood from our definition of the effective temperature, which
includes the extra luminosity of $^{56}$Ni decay above the photosphere. 

Unlike luminosity and temperature, the photospheric velocity is almost
unaffected by  the existence of external
  $^{56}$Ni. Therefore, we do not present a comparison between the
velocities of the optimal model and observations because this is
essentially the same that we presented in
\S~\ref{sec:CT09} and Figure~\ref{fig:velo}. 

One could expect that the presence of $^{56}$Ni and associated
iron-group elements in the outer layers can introduce lines and
increase the blanketing, thus affecting the observed
spectrum. However, if the iron-rich material is confined to a small
solid angle, as in the case of a jet, the effect will be diluted, as
explained by \citet{2012MNRAS.424.2139D}. 

  As our simulations are carried out in one dimension we do not
  have the ability of reproducing the actual structure of a jet-like
  distribution. Nevertheless, in the proposed scenario only a small
  fraction of the mass is involved in the jet itself, thus not
  affecting the global explosion dynamics. The low level of {\em continuum}
  polarization found by \citep{2009ApJ...705.1139M} indicates that
  departures from spherical symmetry should be small. This allows us
  to treat the problem approximately in spherical
  symmetry. Calculations performed in higher dimensions are
  required to produce a self-consistent model that would allow to test
  the proposed jet and aspherical $^{56}$Ni distribution. 

\section{ALTERNATIVE APPROACHES}
\label{sec:discussion}
\subsection{Cooling of the Shocked Envelope}
\label{sec:cooling}
In the literature, the early emission of the LC of SN~2008D was explained 
as a consequence of the cooling of the outer stellar envelope
following the passage the shock through the star and its subsequent
breakout. Analytic models by \citet{waxman07} (W07) and
\citet{2008ApJ...683L.135C} (CF08) were used in
\citet{2008Natur.453..469S} and \citet{2009ApJ...702..226M} to compare
with the early data of SN~2008D. These models describe the emission
of the outer layer of the ejecta assuming: (1) pre-explosion density
$\rho \, \propto \, (1-r/R)^n$, which is valid while the
  photosphere is in the outer shock-accelerated part of the ejecta
  as long as the mass above of the photosphere
  is less than 0.1 $M_\odot$, (2) self-similar solution 
once the supernova reached the state of free expansion to determine
the  post breakout density and velocity, and (3) constant opacity. The
luminosity was calculated in a different way in each model. While W07
did not take into account the radiative diffusion
assuming a strictly adiabatic expansion, CF08 considered the motion of
a diffusion wave through the ejecta. Therefore these models are valid
after free expansion is achieved and until more or less the onset of
recombination, when the 
photosphere begins to recede into the ejecta. Recently, 
\citet{2011ApJ...728...63R} (RW11) improved the model of W07 by
including the effect of recombination on the opacity. They also
corrected a typographical error found in a coefficient of the
equation (19) of W07. 

The analytic models mentioned above provide expressions for the time
evolution of the luminosity, photospheric radius and temperature as a
function of ejecta mass ($M_{\mathrm{ej}}$), ejecta kinetic energy ($E_K$) and
progenitor radius ($R$), besides other parameters that depend on the
structure of the progenitor. However, the dependence of luminosity
and temperature on $R$ is stronger than on the other parameters. It is
linear for the luminosity and $\propto \, R^{1/4}$ for the temperature. 

\begin{figure} 
\begin{center}
\includegraphics[scale=.33, angle=-90]{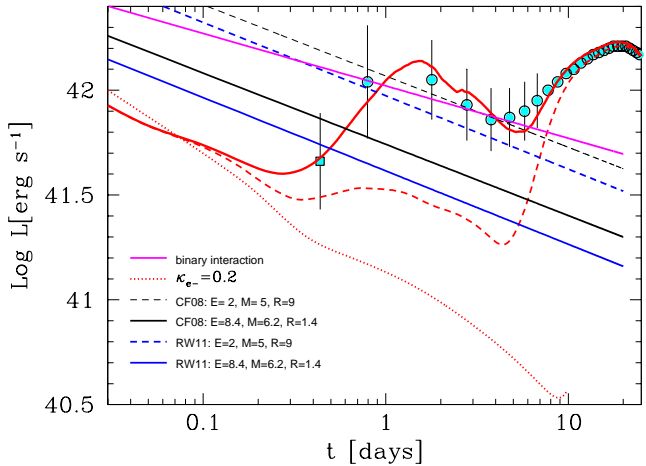}
\caption{Comparison between the bolometric luminosity of the analytic and
  hydrodynamical models. The data are shown with cyan dots. Three
  different hydrodynamical calculations 
  are shown: the optimal model (solid line), the same model but
  without external $^{56}$Ni (dashed line), and a model assuming a
  constant opacity equal to the electron-scattering opacity for
  pure helium material, $\kappa_{e^{-}}= 0.2$ g cm$^{-3}$, (dotted
  line). The analytic models of CF08 (black) and RW11 (blue) for the
  same explosion 
  parameters as in our hydrodynamic simulations are also shown
  (solid line). With dashed
  lines we show the same analytic models for 
  an alternative set of physical parameters that include a larger
  initial radius. Note that a larger radius provides a better fit to
  the observations, with the exception of the earliest
  point. The analytic prediction of the binary interaction model 
   of \citet{2010ApJ...708.1025K} is also included (see
   \S~\ref{sec:binary} for additional 
  information).\label{fig:LCA}} 
\end{center}
\end{figure}

In Figure~\ref{fig:LCA} we show a comparison between our
hydrodynamical calculations and the analytic predictions of CF08 and
RW11. Three different hydrodynamical calculations are shown: the
optimal model, the same model without external $^{56}$Ni, and a model
with constant opacity equal to the electron-scattering opacity for
pure helium, $\kappa_{e^{-}}= 0.2$ g cm$^{-3}$. For the
analytic models, the corrected expression for constant opacity from
RW11 [their equation (13)] was used here instead of equation (19) of
W07. Two sets of explosion parameters for models CF08 and RW11 are
shown. First, we focus on the behavior of the analytic prediction with
the same explosion parameters as in our hydrodynamic simulations
(i.e., $E_K = 8.4$ foe, $M_{\mathrm{ej}}=6.2$ $M_\odot$, and $R= 1.4
R_\odot$).   The analytic models for compact stellar structures
  assume an initial density profile with index $n=3$, which is similar
  to the shape of the initial structure from stellar
  evolution that we adopted for model He8, as shown in the inset of
  Figure~\ref{fig:rho_radii}. As the proportionality constant between
  $\rho$ and $r$ is not given in the analytic models, we have scaled
  the density of these models to match the value of the He8 model
  at $0.1$ $M_\odot$ inside the star, i.e. where the
  analytic models are valid.

From Figure~\ref{fig:LCA} it is clear that (1) the analytics models
predict a higher luminosity than the hydrodynamic model without
external $^{56}$Ni, but they are still not consistent with the
observations, and (2) the slope of the analytic and hydro models are
quite similar until $t \approx 0.5$ d, when helium starts to recombine
and the luminosity of the hydro model enters a nearly plateau phase
for approximately five days. After that time, the heating by $^{56}$Ni
produces a re-brightening of the LC.  
Note that the presence of a post-breakout plateau for Wolf-Rayet
progenitors of different radii and masses has recently been
reported in simulations by \citet{2011MNRAS.414.2985D}. In agreement
with our simulations, they found that the typical differences between
the luminosity of the post-breakout plateau ($1-5 \times 10^7
L_\odot$) and the main peak are much larger than the observed value of
$\approx \, 0.3$ dex for SN~2008D. 
Our hydrodynamical models show that the hypothesis of constant
opacity breaks at $t \approx 0.5$ d, when He begins to recombine
and the photosphere recedes into the ejecta. By $t \approx 1.5$ d, 
the mass above of the photosphere is larger than $0.1$ $M_\odot$ and the
assumptions of the analytical models are no longer valid. Therefore,
our simulations establish a limit of about 
$1.5$ days for the validity of the analytic expressions. 
 This is in close concordance with the range of validity suggested by
 equations (16) 
and (17) of RW11, which for the values of $E_K$, $M_{\mathrm{ej}}$ and 
$R$ used here, give a range of up to $2$ days. Note that for this
  range of time there are only two data points available,
  including the earliest point
    which seems not to follow the analytic predictions. This makes 
  the conclusions derived from the analytic models more dubious. 

Finally, as we noted before, there is a difference in luminosity
  between our models and the analytic models even for times earlier than
  $0.5$ day. The differences may be related with the density structure
  of the outermost layers of the ejecta used in each calculation.
    A direct comparison of initial density structures is not possible
    because the scale of the relation between $\rho$ and $r$ is not
    provided in the analytic calculations. We can however make a
    quantitative comparison of post-shock breakout density
    structures. For the analytic models, this has the form $\rho \,
    \propto \, v^{n}$. Equation~(1) of CF08 evaluated at $E= 8.4$ foe,
    $M= 6.4$ $M_\odot$, and $t= 1.1$ days gives  $\rho = 7.9
      \times 10^{3} \, v^{-10.18}$ g cm$^{-3}$, where $v$ is
    expressed in $10^8$ cm s$^{-1}$. For  
our density profile at $t=1.1$ days we find a similar exponent but a
very different proportionality constant:  $\rho = 1.9  \times
  10^{2} \, v^{-9.66}$ g 
cm$^{-3}$. This difference can be the reason for the discrepancy
  in luminosity between analytic and numerical models. The difference
  in density may occur because the calibration of the post-explosion
density profile used in CF08 was based on previous hydrodynamical
calculations applied to a blue supergiant structure, useful to model
SN~1987A, instead of a Wolf-Rayet progenitor as required here.

\subsection{Different Progenitor Structure} 
A larger value of the progenitor radius of 9 $R_\odot$ was suggested
by CF08 in order to explain the XRT of SN~2008D as completely thermal
emission. On the other hand, \citet{2009ApJ...702..226M} found a
similar radius by fitting the CF08 relations to the early data, and
assuming $E_K = 2$ foe and $M_{\mathrm{ej}}= 5$ $M_\odot$, as suggested by
\citet{2008Natur.453..469S}. In Figure~\ref{fig:LCA}, we include the
analytic models for these alternative physical parameters. In this
case, the analytic models reproduce better the early luminosity with
the exception of the earliest data point.  

The improvement in the fit of the analytic models with larger radius
suggests that adopting progenitors with large radii in our
hydrodynamical simulations may allow us to reproduce
the early data {\em without resourcing to external $^{56}$Ni}. We
thus attached several envelopes in hydrostatic
and thermodynamical equilibrium to the He-rich layer of our He8
model.  The attached envelopes were integrated inward assuming a
  fixed stellar radius and varying the effective temperature and
  envelope mass so that the condition of continuity at the fitting
  point was achieved for mass, luminosity, pressure, and temperature.
We also verified that the bottom of the envelope is cool enough to
prevent nuclear reactions from developing in that 
region. This allowed us to generate a set of progenitors with radii
of 9, 20, 50 and 100 $R_{\odot}$. 

\begin{figure}
\begin{center}
\includegraphics[scale=.33,angle=-90]{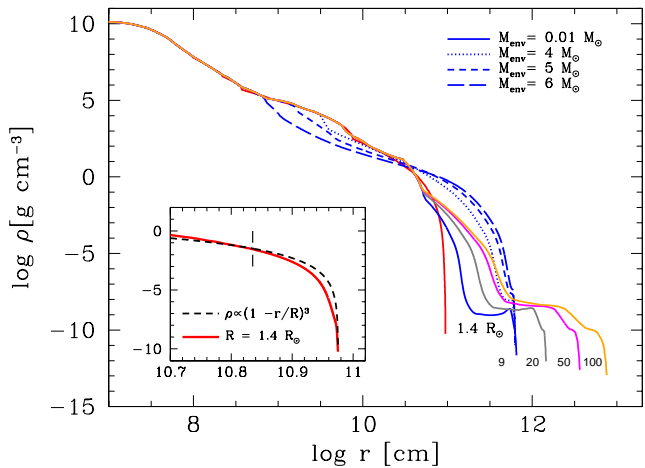}
\caption{Initial density distributions as a function of
  radius for model He8 ($R= 1.4 \, R_\odot$; red line) and for models
  with different initial radii. Thick lines represent models whose
  variations in radius are accomplished by attaching essentially
  mass-less ($<0.01$ $M_\odot$) envelopes to the He-rich layer of He8
  model while thin lines show models with 9 $R_\odot$ and 
  massive envelopes attached at different points of mass inside the
  He8 model, as indicated in the upper right part of the
  figure.  {\em Inset:} blow out of the outermost layers for model He8
  ($R= 1.4 \, R_\odot$; red line) compared with the shape assumed for
  the analytic models: $\rho \propto (1-r/R)^n$, with $n=3$ and $R=
  1.4 \, R_\odot$ (dashed black line). The location of the limit of validity
  of the analytic models ($0.1$ $M_\odot$ inside the star) is
  indicated with vertical lines. \label{fig:rho_radii}}      
\end{center}
\end{figure}

\begin{figure}
\begin{center}
\includegraphics[scale=.33, angle=-90]{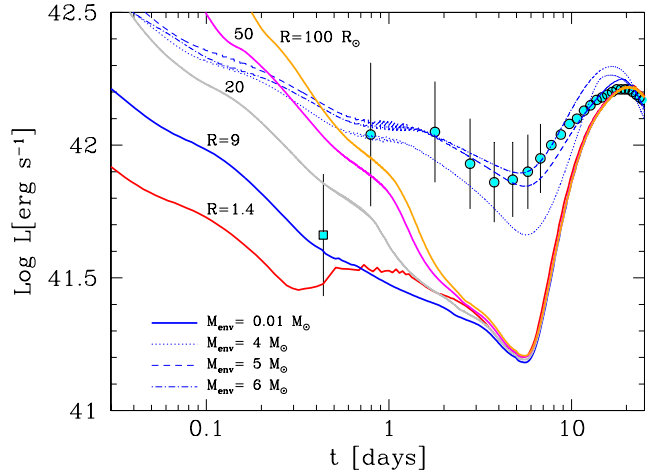}
\caption{Bolometric LC for models with the same explosion energy as
  the optimal model, but different initial radii. Models whose
  variations in radius are accomplished by attaching essentially
  mass-less ($<0.01$ $M_\odot$) envelopes to the  He-rich layer of He8
  model are shown with thick lines. Note that in this case, larger
  radii produce higher early luminosity for $t<6$ days, although not
  high enough to explain 
  the first peak of SN~2008D. Thin lines represent models with 9
  $R_\odot$ but with massive envelopes attached at different points
  of mass inside the He8 model, as indicated in the labels of the
  figure. 
  These models, especially $M_{\mathrm{env}}=$ 5 and 6 $M_\odot$ give a
  reasonable fit to the early observations with the exception of the
  earliest data point.\label{fig:LCRADI}}  
\end{center}
\end{figure}

The resulting bolometric LCs are shown 
with thick lines in Figure~\ref{fig:LCRADI}, where the other
physical parameters are the same as previously adopted. From the 
figure it is clear that at $t\lesssim 5$ days models with larger radii
  produce slower cooling of the outer layers and 
higher luminosity, though not as large as the one predicted by the
  analytical models. After that time, all models look 
remarkably similar. Note that models start to converge 
at $t\approx 1.5$ days when the analytic expressions become invalid.
Finally we see that even with a radius as large as 100 
$R_{\odot}$, we could not satisfactorily reproduce the early LC.

 The analysis above was performed using nearly mass-less envelopes
($<0.01$ $M_\odot$).  The initial density profiles as a function of
radius for these models are shown with thick lines in
Figure~\ref{fig:rho_radii}. The shape of 
the density in the outermost layers not to follow the one used
for the analytical models. This is reflected in a different, smoother, 
dependence of the luminosity on the progenitor radius than the
prediction of the analytic models.

\begin{figure}
\begin{center}
\includegraphics[scale=.33,angle=-90]{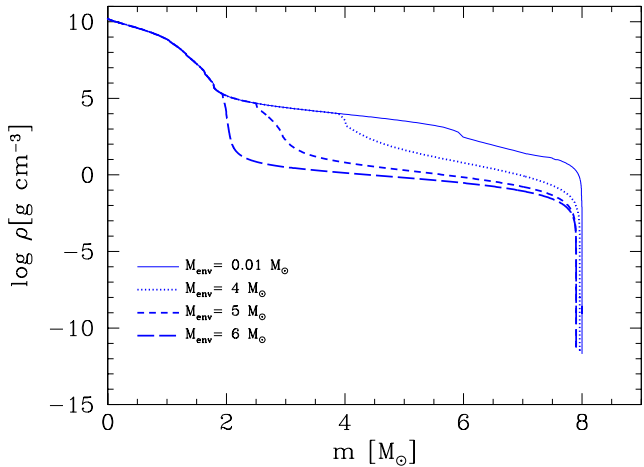}
\caption{Initial density distributions as a function of
  mass for models with  9 $R_\odot$ formed from model He8 but
 with massive envelopes attached at different points of mass, as
 indicated in the labels.\label{fig:rho_mass}} 
\end{center}
\end{figure}

 Alternatively, we calculated initial models with 
9 $R_{\odot}$ but with a substantial modification of the initial
density distribution with respect to the He8 model. Specifically, we
removed  4, 5 and 6 $M_\odot$ of the original He8 model and attached 
new massive envelopes preserving the total mass (8
$M_\odot$), out to a radius of 9 $R_{\odot}$. These structures 
are shown in Figure~\ref{fig:rho_mass} as a function of mass. The
mass-less envelope model for 9 $R_{\odot}$ is also included for
comparison. Figure~\ref{fig:LCRADI} shows the LCs
resulting from these models (thin lines). The early observations can
be reproduced reasonably well with 
these type of structures, excluding the earliest data point. In these
cases, the shape of the density profile as
a function of radius in the outermost layers
(thin lines in Figure~\ref{fig:rho_radii}) are closer to the function
assumed for the analytic models.

  This means that, without considering the earliest data point, models
  with the larger radius and a modified density structure than those
  predicted by stellar evolution calculations can give a reasonable
  explanation of the early LC of SN~2008D as the cooling expansion of
  the outer envelope. However, the massive-envelope models give a
    much poorer fit to the LC around the main peak, and they pose
    the additional problem of finding a physical explanation to justify
  a density profile that is 
  different from those predicted by stellar evolution models.
  Note that some variations in the assumed pre-explosion
    density profile can be caused by effects of rotation, which were not 
    included in our initial models. Nevertheless, the expected
    differences are much smaller than those needed to fit 
    the early data of SN~2008D. 

  It is important remark 
  that recent binary stellar evolution calculation presented by
  \citet{2010ApJ...725..940Y} predicted a larger progenitor radius
  than that of $1.4$ $R_\odot$ obtained for our He8 model based on
  single stellar evolution. However, the radius comprising 95\% of the
  mass is less than 1
  $R_\odot$, with the exception of the models that contain some hydrogen
  for which this radius can be as large as 5 $R_\odot$ \citep[see Table~2
    of][]{2010ApJ...725..940Y}. For SN~2008D, the
  presence of a thin H envelope was firmly ruled out from
  spectroscopic analysis, e.g. T09 estimated a very low upper limit
  for the hydrogen mass fraction of $5 \times 10^{-4}$
  $M_\odot$. Therefore, the binary models are similar to our low-mass 
  envelope models and are not consistent with the early
  observations of SN~2008D. This also becomes clear from the results
  presented by \citet{2011MNRAS.414.2985D} based on the binary models of
  \citet{2010ApJ...725..940Y}. All the LCs presented by
  \citet{2011MNRAS.414.2985D} have a post-breakout luminosity that is much
  smaller than the one observed for SN~2008D,
 as noted by the authors and according also to our calculations.

\subsection{Binary interaction model} 
\label{sec:binary}
As we noted previously, \citet{2011MNRAS.414.2985D} also found that the 
observed luminosity of SN~2008D at early times was much larger than
the one estimated by their models. Their
calculations were done with a different hydrodynamical code
and evolutionary initial models, as compared with our
study. They considered unlikely that light contamination from the
host galaxy, or any light scattered by the CSM or pre-SN mass loss
could produce the large post-breakout luminosity. Moreover,
CF08 estimated that the mass-loss rate of the progenitor was too low
for the CSM to be optically thick. Alternatively,
 \citet{2011MNRAS.414.2985D} 
 suggested that the large observed luminosity could be due to the
collision of the SN ejecta with a companion star in
a binary system. However, they did
not perform any detailed test of this scenario. Models that explain
the early enhancement of the luminosity due to binary
collision were proposed by \citet{2010ApJ...708.1025K}. These models
depend on several parameters, such as binary separation, mass
of the ejecta, shock velocity and viewing angle.
An analytic expression for the luminosity as a function of time
for a viewing angle of 45$^\circ$ is given in Equation~(22) of
\citet{2010ApJ...708.1025K}, with a 
stronger dependence on the shock velocity and the binary separation
($a$) than on the ejected mass and the electron opacity.
The shock velocity for SN~2008D can be estimated from the ejecta mass and explosion
energy, assuming that $\kappa_{e^{-}}= 0.2$ g cm$^{-3}$
for fully ionized electron-scattering of pure
helium, and thus leaving $a$ as the only free parameter of the
problem. Figure~\ref{fig:LCA} shows an example of the binary
collision model for a shock
  velocity derived using an explosion energy of $8.4$ foe and an ejecta mass
of $6.2$ $M_\odot$ as adopted in this paper.
For this set of
parameters the orbital separation required to reproduce the early
luminosity is $a=1.8 \, R_\odot$. This value can be slightly increased
to $a=2.8  \,R_\odot$ if we assume  values of  $E=3.7$ foe and 
 $M_{\mathrm{ej}}= 4.4 \;M_\odot$ appropriate for the He6 model (see
section~\ref{sec:CT09}).
This very small constraint found for the
binary separation may pose a problem for the binary collision
scenario. Note that the typical binary
separation of Galactic Wolf-Rayet stars is $\gtrsim 10 R_\odot$, 
with very few exceptions \citep{2001NewAR..45..135V}. 
The situation might be remedied by 
assuming smaller viewing angles \citep[see Figure~2
  of][]{2010ApJ...708.1025K}, but since we have no way to test this
scenario, we will consider it a less likely possibility.

\section{CONCLUSIONS}
\label{sec:conclusion}
The early post-breakout emission of SN~2008D cannot be
explained as the cooling of the outer layers of a normal Wolf-Rayet
star after the passage of the shock wave, as suggested in the
literature. Alternatively we have proposed a model which 
assumes a double-peaked $^{56}$Ni distribution to explain this early
emission as an attractive solution to this problem. Specifically,
an amount of $0.01 \, M_\odot$ of $^{56}$Ni was located in the outermost
layers of the ejecta with $v \gtrsim  20,000$ km s$^{-1}$---although the exact
amount and distribution is subject to some uncertainties inherent to
the data and the model itself. The assumption of external $^{56}$Ni
allowed us to reproduce very well both the early and late observations of
SN~2008D. The presence of this high-velocity
radioactive material may be caused by the formation of jets during the
explosion. A multidimensional model would be required to confirm
  our suggestion in a self-consistent fashion.

Hydrodynamical models applied to Wolf-Rayet star predict a much larger
contrast ($\gtrsim 0.9$ dex) between the initial dip and the peak of
the LC due to heating by $^{56}$Ni than what is observed for SN~2008D
($\approx$ $0.3$ dex). Only with a substantial modification of 
the initial density structure that is predicted by stellar
evolution calculations and for a larger progenitor radius, 9 R$_\odot$,
can the early emission be compatible with the cooling of the outer
envelope. Even in this case the earliest observed point is not
reproduced by the shock-cooling model and a much worse fit to
  the LC around the main peak is obtained. In addition, some physical 
explanation for such substantial changes in the initial density is
needed to consider this as a possibility. 

We also analyzed the possibility that the early emission was due to 
  interaction of the ejecta with a binary companion. This
  was based on the analytic 
  predictions for the luminosity given by \citet{2010ApJ...708.1025K}. 
However, we found that the binary separation required to explain the early
luminosity of SN~2008D is very small ($\lesssim 3 \,R_\odot$), which
poses a serious conflict for this interpretation.

Furthermore, we note that only the double-peaked $^{56}$Ni model can 
  reproduce the earliest observed data point as well as
  the rest of the LC. Both the binary interaction model and the modified
  density structure predict a larger luminosity at the earliest observed
  epoch.

Comparing our hydrodynamical models with the analytic models commonly
adopted in the literature to reproduce the early emission (e.g.; CF08
and RW11) we found: 
(1) the range of applicability of the analytic formula is
very restricted. An upper limit of 
 $\approx 1.5$ days after the explosion was found. 
In this range of
time only two data points of SN~2008D are available. (2) The analytic models
were calibrated with structures that may not apply to
Wolf-Rayet stars. This could be the reason for the  
very different value found for the proportionality constant of the
relationship $\rho \propto v^{-n}$ for the post-explosion density
profile as compared with the one used in CF08 (Equation~1). (3) A 
modification of the radius of the star, as commonly applied for the analytic
models, implies a different inner boundary 
condition for the density which cannot represent any stellar
evolutionary solution. Therefore, conclusions based on these simple
models should be taken with caution.

The analysis presented here shows the relevance of the early emission
in our understanding of the progenitor structure previous to 
the explosion. We expect that current SN searches of increasing
cadence will frequently detect more CCSNe in the early stages of their
evolution. Then it will be possible to test whether the early behavior
of SN~2008D is a peculiar to it or if it is a common feature of some
type of SNe.

\acknowledgments
The authors gratefully acknowledges the helpful conversations with
Keiichi Maeda and Gaston Folatelli. This research has been supported
in part by the Grant-in-Aid for Scientific Research of MEXT (22012003
and 23105705) and JSPS (23540262) and by World Premier International
Research Center Initiative, MEXT, Japan.

\end{document}